\documentclass[11pt]{article}
\usepackage{graphicx}
\begin{document}
{\setlength{\oddsidemargin}{1.2in}
\setlength{\evensidemargin}{1.2in} } \baselineskip 0.55cm
\begin{center}
{{Quantum gravity effects on Hawking radiation of Schwarzschild-de Sitter black holes}}
\end{center}
\begin{center}
  ${\rm T.\, Ibungochouba\, Singh}^{1}$, ${\rm I.\, Ablu\, Meitei}^{2}$, ${\rm K.\, Yugindro\, Singh}^{3}$ \\
1: Department of Mathematics, Manipur University, Canchipur, Manipur (India)\\
2: Department of Physics, Modern College, Imphal, Manipur (India)\\
3: Department of Physics, Manipur University, Canchipur, Manipur (India)\\
1: E-mail: ibungochouba@rediffmail.com\\
2:E-mail: ablu.irom@gmail.com
\end{center}
\date{}

\begin{abstract}
The correction of Hawking temperature of Schwarzschild-de Sitter (SdS) black hole is investigated using the generalized Klein-Gordon equation and the generalized Dirac equation by taking the quantum gravity effects into account. We derive the corrected Hawking temperatures for scalar particles and fermions crossing the event horizon. The quantum gravity effects prevent the rise of temperature in the SdS black hole. Besides correction of Hawking temperature, the Hawking radiation of SdS black hole is also investigated using massive particles tunneling method. By considering self gravitation effect of the emitted particles and the space time background to be dynamical, it is also shown that the tunneling rate is related to the change of Bekenstein-Hawking entropy and small correction term $(1+2\beta m^2)$. If the energy and the angular momentum are taken to be conserved, the derived emission spectrum deviates from the pure thermal spectrum. This result gives a correction to the Hawking radiation and is also in agreement with the result of Parikh and Wilczek.

\end{abstract}

1. {\bf Introduction} \setcounter {equation}{0}
\renewcommand {\theequation}{1.\arabic{equation}}

Hawking showed that black hole radiates like a black body radiation [1-2]. After that different methods have been proposed for deriving Hawking radiation. Parikh and Wilczek [3] investigated the tunneling behavior of massless scalar particles by considering the background variation in black hole evaporation, which is known as null geodesic method. Later, Angheben et al. [4] investigated the Hawking radiation as tunneling by using Hamilton-Jacobi method and WKB approximation. The importance of applying Hamilton-Jacobi method is using WKB approximation and Feynman prescription to calculate the imaginary part of the action for the tunneling process. Following this method, many works have been done [5-9]. Recently, tunneling of Dirac particle through the event horizon was investigated by Kerner and Mann for the Rindler space time and general rotating black hole space time. In this process, using Pauli Sigma matrices and substituting appropriate wave function into Dirac equation, the action of the radiant Dirac particle can be obtained which, in turn, is related to the Boltzmann factor for emission at Hawking temperature according to semiclassical WKB approximation [10]. Using their method, the Hawking temperatures were obtained for fermions tunneling complicated black hole space times [11-14]

The different theories of quantum gravity such as string theory, loop quantum gravity and quantum geometry prove the existence of a minimal length [15-19]. This minimal length can be obtained using generalized uncertainty principle (GUP) through modified commutation relation. The fundamental commutation relation must be modified so as to derive GUP. There are two modified commutation relations. Firstly, Kempt et al. [20] put forward the expression of GUP as
$\Delta x \Delta p\geq\frac{\hbar}{2}[1+\beta(\Delta p)^2]$, where $\beta=\beta_0\frac{\ell^2_p}{\hbar^2}$. $\beta_0<10^5$ and $\ell_p$ are dimensionless parameter and the Plank length respectively. The modified commutation relation is defined by $[x_a, p_b]=i\hbar\delta_{ab}[1+\beta p^2]$, where $x_a$ and $ p_b$ are the position and momentum operators. Also $x_a$ and $ p_b$ are defined by  $x_a=x_{0a}$ and $p_b=p_{0b}[1+\beta p^2]$. $x_{0a}$ and $p_{0b}$ should satisfy the canonical commutation relations $[x_{0a}, p_{0b}]=i\hbar\delta_{ab}$. Secondly, Das, et al. [21-22] extended the GUP based on doubly special relativity known as DSR-GUP.

Applying modified form of GUP, Majhi and Vagenas [23] investigated the Unruh effect. Based on the modification, Kim et al. [24] and Xiang et al. [25] discussed black hole thermodynamics and the tunneling rate was investigated in [26-27]. The creation of scalar particle pairs by an electric field in the presence of a minimal length was investigated by Haouat and Nouicer [28]. Taking the quantum gravity into account influenced by DSR-GUP and Parikh-Wilczek tunneling method, Nozari and Saghafi [29] investigated the radiation of massless scalar particles in the schwarzschild black hole. Recently, Chen et al. [30] studied the fermion tunneling effect for the Schwarzschild black hole using the generalized Dirac equation. They showed the remnant in black hole evaporation is $\geq\frac{M_{p}}{\beta_0}$, where $M_{p}$  is the Plank mass. Following their works, more fruitful results have been obtained [31-35].

The aim of this paper is to investigate the tunneling behavior of scalar particles and fermions across the black hole event horizon of a SdS black hole taking the quantum gravity effects into account. Using generalized Klein-Gordon equation and generalized Dirac equation, the corrections of Hawking temperature are recovered. By taking space time background to be dynamical and self-gravitational effect into account, we show that the tunneling of scalar particles is related to the change of Bekenstein-Hawking entropy and  the term $(1+2\beta m^2)$, where $m$ is the mass of the particle.

The organization of this paper is as follows. In Sect. 2, by considering the quantum gravity effect into account, the correction to the Hawking temperature of SdS black hole is studied using generalized Klein-Gordon equation and WKB approximation. In Sect. 3, applying modified Dirac equation, the tunneling radiation of fermions in the SdS black hole is discussed. Section 4 is the conclusion of our paper.

 2. {\bf Schwarzschild-de Sitter black hole and tunneling of scalar particles}. \setcounter {equation}{0}
\renewcommand {\theequation}{\arabic{equation}}
The line element describing Schwarzschild-de Sitter black hole in
Boyer-Lindguist coordinates [8] is given by
\begin{eqnarray}
ds^2&=&-\Big(1-\frac{2M}{r}-\frac{r^2}{\ell^2}\Big)dt^2+\Big(1-\frac{2M}{r}-\frac{r^2}{\ell^2}\Big)dr^2+r^2(d\theta^2+r^2\sin^2\theta d\phi^{2}),
\end{eqnarray}
where $\ell^2=\frac{3}{\Lambda}$ and $M$ is the mass of the black hole. As $\ell$ tends to infinity, then the Eq. (1) represents Schwarzschild black hole and the space time metric is type D in the Petrov classification. At the large value of $r$, the given metric becomes dS space, in such case the space time metric is conformally flat. Also the given metric will represent anti SdS black hole if $-\ell^2$ is replaced by $\ell^2$. The SdS black hole has real positive singularity when $\frac{1}{\ell^{2}r}(r-r_{h})(r-r_{c})(r_{-}-r)=0$. The locations of the event horizon, $r_{h}$ and the cosmological horizon, $r_{c}$ are given by
\begin{eqnarray}
r_{h}=\frac{2M}{\sqrt{3\Xi}}\cos\frac{\pi+\Phi}{3},\cr
r_{c}=\frac{2M}{\sqrt{3\Xi}}\cos\frac{\pi-\Phi}{3},
\end{eqnarray}
where $\Phi=\cos^{-1}(3\sqrt{3\Xi})$ and $\Xi=\frac{M^{2}}{\ell^{2}}$. Expanding the event horizon in terms of mass of the black hole with $\Phi<\frac{1}{27}$, we get
\begin{eqnarray}
r_{h}=2M(1+\frac{4M^{2}}{\ell^{2}}+\cdot\cdot\cdot)\cdot
\end{eqnarray}
In order to study particle tunneling from the SdS black hole near the event horizon, the SdS metric can be written as
\begin{eqnarray}
ds^2&=&-F(r)dt^2+\frac{1}{G(r)}dr^2+g_{\theta\theta}d\theta^2+g_{\phi\phi}d\phi^{2},
\end{eqnarray}
where $F(r)=G(r)=\frac{\Delta_{,r}(r_{h})(r-r_{h})}{r^{2}}$ and $\Delta_{,r}(r_{h})=2(r_{h}-M-\frac{2r^{3}_{h}}{\ell^2})$.
The generalized form of Klein-Gordon equation is given by [31]
\begin{equation}
-(i\hbar)^{2}\partial^{t}\partial_{t}\Psi=\Big[(i\hbar)^{2}\partial^{t}\partial_{t}+m^2\Big]\Big\{1-2\beta\Big[(i\hbar)^{2}\partial^{t}\partial_{t}+m^2\Big]\Big\}\Psi.
\end{equation}
For studying the equation of motion, the wave function of the scalar particle can be taken as
\begin{equation}
\Psi={\rm exp}\Big[\frac{i}{\hbar}I(t, r, \theta, \phi)\Big].
\end{equation}
Using Eqs. (4) and (6) in Eq. (5), we get
\begin{eqnarray}
\frac{1}{F}(\partial_{t}I)^{2}&=&\Big[G(\partial_{r}I )^{2}+\frac{1}{r^2}(\partial_{\theta}I)^{2}+\frac{1}{r^2\sin^2\theta}(\partial_{\phi})^{2}+m^2\Big]\cr&&
\times\Big\{1-2\beta\Big[G(\partial_{r}I )^{2}+\frac{1}{r^2}(\partial_{\theta}I)^{2}+\frac{1}{r^2\sin^2\theta}(\partial_{\phi})^{2}+m^2\Big].
\end{eqnarray}
It is very difficult to solve Eq. (7) because it contains the variables $t, r, \theta, \phi$. For studying Hawking radiation near the black hole by taking quantum gravity effect into account, the action $I$ can be taken as
\begin{equation}
I=-\omega t+R(r)+ W(\theta, \phi),
\end{equation}
where $\omega$ is the energy of the emitted scalar particle. We assume that the particle tunnels along the radial direction,
then we get
\begin{equation}
\frac{1}{r^{2}}(\partial_{\theta}W)^{2}+\frac{1}{r^2\sin^2\theta}(\partial_{\phi}W)^{2}=e.
\end{equation}
In the above equation, $e$ is a constant and can be made as zero. Then Eq. (7) becomes a biquadratic equation as follow
\begin{equation}
a(\partial_{r}R)^{4}+b(\partial_{r}R)^{2}+c=0,
\end{equation}
where
\begin{eqnarray}
a&=& -2\beta G^{2}(r),\cr
b&=& G(1-4\beta e-4 \beta m^2),\cr
c&=& m^2+e-2\beta e^2-4\beta e m^2-2\beta m^4-\frac{\omega^2}{F(r)}.
\end{eqnarray}
We assume that $b^{2}-4ac=G^2(1-8\beta\frac{\omega^2}{F(r)})>0$. Eq. (10) has four roots. The two roots having physical meaning at the event horizon of black are given by
\begin{eqnarray}
R_{\pm} &=&\pm \int \frac{1}{\sqrt{FG}}\sqrt{\omega^{2}-m^{2}F+2\beta m^{4}F}\Big[1+2\beta m^{2}\Big]dr,\cr
&=&\frac{4i\pi M^{2}\omega(1+\frac{4M^2}{\ell^2}+\cdot\cdot\cdot)^2}{[2M(1+\frac{4M^2}{\ell^2}+\cdot\cdot\cdot)-M-\frac{16M^2}{\ell^2}(1+\frac{4M^2}{\ell^2}+\cdot\cdot\cdot)^3]}
(1+2\beta m^{2}).
\end{eqnarray}
Then the imaginary part of the action can be written as
\begin{eqnarray}
{\rm Im} R &=&\pm\frac{4\pi M\omega(1+\frac{8M^2}{\ell^2})}{(1-\frac{8M^2}{\ell^2})}
(1+2\beta m^{2}),
\end{eqnarray}
where the terms $M^{n}$ for $n\geq5$ is neglected. Also $\pm$ sign represents outgoing/ingoing wave and $F=G=\frac{\Delta_{,r}(r_{h})(r-r_{h})}{r^{2}}$. The tunneling rate of the scalar particles crossing the black hole event horizon is
\begin{eqnarray}
\Gamma&=&\frac{P_{({\rm emission})}}{P_{({\rm absorbtion})}}=\frac{exp(-ImI_{+})}{exp(-ImI_{-})}\cr
&=&\frac{exp(-ImR_{+}-ImW)}{exp(-ImR_{-}-ImW)}\cr
&=&exp\Big[-\frac{8\pi M\omega(1+\frac{8M^2}{\ell^2})}{(1-\frac{8M^2}{\ell^2})}
(1+2\beta m^{2})\Big].
\end{eqnarray}
 The Boltzmann factor with Hawking temperature is given by
\begin{eqnarray}
T=\frac{(1-\frac{8M^2}{\ell^2})}{8\pi M\omega(1+\frac{8M^2}{\ell^2})(1+2\beta m^{2})} =\frac{T_{0}}{(1+2\beta m^{2})},
\end{eqnarray}
where $T_{0}=\frac{(1-\frac{8M^2}{\ell^2})}{8\pi M\omega(1+\frac{8M^2}{\ell^2})}$ is the original Hawking temperature of SdS black hole. It indicates that the correction of Hawking temperature can be observed for the small value of $\beta$. From the direct calculation, the corrected value of Hawking temperature depend on the mass of the black hole and mass of the emitted scalar particles but not explicitly on the energy of the emitted particles. It is also observed that the corrected Hawking temperature under the consideration of quantum gravity effect is lower than the original one. This implies that the quantum gravity effect will slow down the rise of Hawking temperature of SdS black hole.

If the small term $\frac{2\beta m^4F}{\omega^2}$ is neglected in the roots of Eq. (10), then Eq. (12) can be written as
\begin{eqnarray}
R &=&\pm \int \frac{1}{\sqrt{FG}}\sqrt{\omega^{2}-m^{2}F+2\beta m^{4}F}\Big[1+\beta (m^{2}+\frac{\omega^2}{F})\Big]dr.
\end{eqnarray}
Then the imaginary part of Eq. (13) can be expressed as follows
\begin{eqnarray}
{\rm Im} R &=&\pm\frac{4\pi M\omega(1+\frac{8M^2}{\ell^2})}{(1-\frac{8M^2}{\ell^2})}
\Big[1+\beta\Big\{ m^{2}+\frac{4\omega^2(1+\frac{12M^2}{\ell^2})(1-\frac{8M^2}{\ell^2})}{(1-\frac{16M^2}{\ell^2})(1+\frac{8M^2}{\ell^2})}\Big\}\Big].
\end{eqnarray}
Similarly, the corrected Hawking temperature (15) can be written as
\begin{eqnarray}
T=\frac{(1-\frac{8M^2}{\ell^2})}{8\pi M\omega(1+\frac{8M^2}{\ell^2})(1+\beta K)} =\frac{T_{0}}{(1+\beta K)},
\end{eqnarray}
where $K=m^{2}+\frac{4\omega^2(1+12M^2\ell^{-2})(1-8M^2\ell^{-2})}{(1-16M^4\ell^{-2})(1+8M^2\ell^{-2})}$. In such case the corrected Hawking temperature not only depend on the mass of the black hole, mass of the emitted scalar particle but also energy of the emitted particles. When $\ell\longrightarrow\infty$, i.e $\Lambda=0$, our result is agreement with the result of Wang et al. [31].

Since we know that the SdS space time is dynamic. By fixing ADM mass of the total space time, the mass of the SdS is allowed to fluctuate. If a particle tunnels out with energy, then the mass of the SdS black hole changed into $M-\omega$. We also observe that the angular velocity of a particle near the horizon and the angular momentum are zero. The self gravitational interaction is taken into account, the imaginary part of Eq. (17) can be expressed in the following integral form
\begin{eqnarray}
{\rm Imp} R &=&-4\pi\int^{M-\omega}_{M}\frac{(M-\omega')(1+\frac{8(M-\omega')^2}{\ell^2})}{(1-\frac{8(M-\omega')^2}{\ell^2})}
(1+2\beta m^{2})d\omega'.
\end{eqnarray}
The tunneling rate of SdS black hole is given by
\begin{eqnarray}
\Gamma\sim exp[\Delta S_{BH}(1+2\beta m^2)]=exp[\pi(r^2_i-r^2_j)(1+2\beta m^2)],
\end{eqnarray}
where $r_i=\sqrt{2(M-\omega)\Big(1+\frac{4(M-\omega)^2}{\ell^2}\Big)}$ and $r_j=\sqrt{2M(1+\frac{4M^2}{\ell^2})}$ denote the location of event horizon before and after
  the emission of the particles from the SdS black hole and $\Delta S_{BH}=\Delta S_{BH}(M-\omega)-\Delta S_{BH}(M)$ is the change of Bekenstein-Hawking entropy.
 If $\beta=0$, our work is in accordance with the result of Rahman and Hossain [8]. When $\ell$ tends to infinity, the pure thermal spectrum of Schwarzschild black hole can be obtained from Eq. (20) as
 \begin{eqnarray}
\Gamma\sim exp[\Delta S_{BH}(1+2\beta m^2)]=exp[-8\pi\omega(M-\frac{\omega}{2})(1+2\beta m^2)],
\end{eqnarray}
 If we set $\beta=0$, Eq. (21) is consistent with the result of Parikh and Wilczek [3].

 3. {\bf Tunneling of fermions}.
For the study of Hawking radiation by taking quantum gravity effects into account, the main problem is to calculate the imaginary part of the action. The Dirac equation in generalized form can be written as
\begin{eqnarray}
&&\Big[i\gamma^{0}\partial_{0}+i\gamma^{i}\partial_{i}(1-\beta m^2)+i\beta \hbar^{2}\gamma^{i}(\partial_{i}\partial^{i})\partial_{i}+\frac{m}{\hbar}(1-\beta m^2+\beta\hbar^{2}\partial_{i}\partial^{i})\cr&&+i\gamma^{\mu}\Omega_{\mu}(1-\beta m^2+\beta\hbar^{2}\partial_{i}\partial^{i})\Big]\Psi=0,
\end{eqnarray}
where $m$ is the mass of the fermion and $\Omega_{\mu}=\frac{i}{2}\omega^{ab}_{\mu}\Sigma_{ab}$, $\omega^{ab}_{\mu}$ is the spin co-efficient defined by tetrad $e^{\lambda}_{b}$
\begin{eqnarray}
\omega^{a}_{\mu\,\,\,\,b}&=&e^{a}_{\nu}e^{\lambda}_{b}\Gamma^{\nu}_{\nu\mu}-e^{\lambda}_{b}\partial_{\mu}e^{a}_{\lambda},\cr
\Sigma_{ab}&=&\frac{i}{4}[\gamma^{a},\,\,\gamma^{b}],\,\,\,\{\gamma^{a},\,\,\gamma^{b}\}=2g^{ab}.
\end{eqnarray}
For this SdS black hole, the component of the $\gamma^\mu$ matrices in spherical coordinate system are defined as

\begin{eqnarray}
\gamma^t&=&\frac{1}{\sqrt{F(r)}}\left( \begin{array}{cc}
 i  & 0   \\
 0  &  -i  \\
 \end{array} \right),\cr\cr \gamma^r&=&\sqrt{G(r)}\left( \begin{array}{cc}
0  & \sigma^3   \\
 \sigma^3  &  0 \\
 \end{array} \right),\cr\cr\gamma^\theta&=&\frac{1}{\sqrt{g_{\theta\theta}}}\left( \begin{array}{cc}
0  & \sigma^1   \\
 \sigma^1  &  0  \\
 \end{array} \right),\cr\cr\gamma^\phi&=&\frac{1}{\sqrt{g_{\phi\phi}}}\left( \begin{array}{cc}
0  & \sigma^2   \\
 \sigma^2  &  0  \\
 \end{array} \right),
  \end{eqnarray}
  where $\sigma^j (j=1,2,3)$'s are the Pauli Sigma matrices. The importance of solving Dirac equation is to find the imaginary part of of the action of the radiant fermion that, in turn, is related to the Boltzmann factor of emission in accordance with the semi classical WKB approximation. For spin up particles, the wave function is defined by
\begin{eqnarray}
\Psi=\left( \begin{array}{c}
0    \\
 X \\
 0    \\
 Y
 \end{array} \right){\rm exp}\Big[\frac{i}{\hbar}I(t,r,\theta,\phi)\Big],
 \end{eqnarray}
where $X$, $Y$ and $I$ are functions of coordinates $t, r, \theta, \phi$ and $I$ is the action of the emitted fermion. In this paper, we will consider spin up case since it is equal to spin down case with some change in sign. Dirac equation can be decoupled only for stationary space time [36] or in the spherically symmetric Vaidya-Bonner black hole [37]. Using Eqs [1], [24], [25] in Eq. [22], the final decoupled equations are obtained as
\begin{eqnarray}
&&-iX\frac{1}{\sqrt{F}}\partial_{t}I-Y(1-\beta m^2)\sqrt{G}\partial_{r}I-Xm\beta\Big[G(\partial_{r}I)^{2}+g^{\theta\theta}(\partial_{\theta}I)^{2}+g^{\phi\phi}(\partial_{\phi}I)^{2}\Big]\cr&&
+Y\beta\sqrt{G}\partial_{r}I\Big[G(\partial_{r}I)^{2}+g^{\theta\theta}(\partial_{\theta}I)^{2}+g^{\phi\phi}(\partial_{\phi}I)^{2}\Big]+Xm(1-\beta m^2)=0
\end{eqnarray}
\begin{eqnarray}
&&iY\frac{1}{\sqrt{F}}\partial_{t}I-X(1-\beta m^2)\sqrt{G}\partial_{r}I-Ym\beta\Big[G(\partial_{r}I)^{2}+g^{\theta\theta}(\partial_{\theta}I)^{2}+g^{\phi\phi}(\partial_{\phi}I)^{2}\Big]\cr&&
+X\beta\sqrt{G}\partial_{r}I\Big[G(\partial_{r}I)^{2}+g^{\theta\theta}(\partial_{\theta}I)^{2}+g^{\phi\phi}(\partial_{\phi}I)^{2}\Big]+Ym(1-\beta m^2)=0
\end{eqnarray}
\begin{eqnarray}
&&X\{-(1-\beta m^2)\sqrt{g^{\theta\theta}}\partial_{\theta}I+\beta\sqrt{g^{\theta\theta}}\partial_{\theta}I[G(\partial_{r}I)^{2}+g^{\theta\theta}(\partial_{\theta}I)^{2}
+g^{\phi\phi}(\partial_{\phi}I)^{2}]\cr&&
-i(1-\beta m^2)\sqrt{g^{\phi\phi}}\partial_{\phi}I+i\beta\sqrt{g^{\phi\phi}}\partial_{\phi}I[G(\partial_{r}I)^{2}
+g^{\theta\theta}(\partial_{\theta}I)^{2}+g^{\phi\phi}(\partial_{\phi}I)^{2}]\}=0\nonumber\\
\end{eqnarray}
\begin{eqnarray}
&&Y\{-(1-\beta m^2)\sqrt{g^{\theta\theta}}\partial_{\theta}I+\beta\sqrt{g^{\theta\theta}}\partial_{\theta}I[G(\partial_{r}I)^{2}+g^{\theta\theta}(\partial_{\theta}I)^{2}
+g^{\phi\phi}(\partial_{\phi}I)^{2}]\cr&&
-i(1-\beta m^2)\sqrt{g^{\phi\phi}}\partial_{\phi}I+i\beta\sqrt{g^{\phi\phi}}\partial_{\phi}I[G(\partial_{r}I)^{2}
+g^{\theta\theta}(\partial_{\theta}I)^{2}+g^{\phi\phi}(\partial_{\phi}I)^{2}]\}=0.\nonumber\\
\end{eqnarray}
In the stationary space time, SdS has time like Killing vector $\frac{\partial}{\partial t}$. In order to obtain radial action, the separation of variables can be performed as
\begin{eqnarray}
I=-\omega t+ W(r)+\Theta(\theta, \phi),
\end{eqnarray}
where $\omega$ is the energy of the emitted fermion. Using Eq (30) into the  Eqs. (28) and (29), we get identical equations after eliminating $X$ and $Y$ as follows
\begin{eqnarray}
&&(\frac{1}{r}\partial_{\theta}\Theta+i\frac{1}{r\sin\theta}\partial_{\phi}\Theta)\cr&&\times[\beta G(\partial_{r}W)^{2}+\beta\frac{1}{r}(\partial_{\theta}\Theta)^{2}+\beta\frac{1}{r^{2}\sin^{2}\theta}(\partial_{\phi}\Theta)^{2}-(1-\beta m^2)]=0.
\end{eqnarray}
We know that $\beta$ is a small quantity as it represents the effect of quantum gravity. In the above equation, the term inside the square brackets can not be equal to zero. From this, we get
\begin{eqnarray}
\frac{1}{r}\partial_{\theta}\Theta+i\frac{1}{r\sin\theta}\partial_{\phi}\Theta=0,
\end{eqnarray}
which gives a complex function (or constant) solution of $\Theta$. It may give rise to a contribution to imaginary part but has no contribution to the tunneling rate. So we can ignore it. Here $g^{\theta\theta}=r^{-2}$ and$g^{\phi\phi}=r^{-2}(\sin\theta)^{-2}$, then we get $r^{-2}\partial_{\theta}\Theta+=r^{-2}(\sin\theta)^{-2}\partial_{\phi}\Theta=0$. Now our aim is to solve the first two equations which can determine the Hawking radiation of the SdS black hole at the event horizon. Using Eq. (32) into Eqs. (26) and (27) and canceling $X$ and $Y$, a polynomial of order six is obtained as
\begin{eqnarray}
L_{6}(\partial_{r}W)^{6}+L_{4}(\partial_{r}W)^{4}+L_{2}(\partial_{r}W)^{2}+L_{0}=0,
\end{eqnarray}
where
\begin{eqnarray}
L_{6}&=&\beta^{2}G^3F,\cr
L_{4}&=&\beta G^2F(m^2\beta+2\beta H-2),\cr
L_{2}&=&GF[(1-\beta m^2)^{2}+\beta(2m^4-2H-2m^4\beta+\beta H^2)],\cr
L_{0}&=&-m^2(1-\beta m^2-\beta H)^{2}F-\omega^2,\cr
H&=&\frac{1}{r}\partial_{\theta}\Theta+i\frac{1}{r\sin\theta}\partial_{\phi}\Theta.
\end{eqnarray}

Using Eq. (32), we have $H=0$ and also ignoring higher order terms of $\beta$, Eqs. (33) and (34) can be written as
\begin{eqnarray}
L_{4}(\partial_{r}W)^{4}+L_{2}(\partial_{r}W)^{2}+L_{0}=0,
\end{eqnarray}
where
\begin{eqnarray}
L_{4}&=&-2\beta G^2F,\cr
L_{2}&=&GF,\cr
L_{0}&=&-\omega^2-m^2F(1-2\beta m^2).
\end{eqnarray}
Eq. (35) has four roots. Solving the roots having physical meaning at the black hole event horizon
\begin{eqnarray}
W_{\pm} &=&\pm \int \frac{1}{\sqrt{FG}}\sqrt{\omega^{2}+m^{2}F}\Big[1+\beta (m^{2}+\frac{\omega^2}{F})\Big]dr,\cr
&=&\pm\frac{4i\pi M^{2}\omega(1+\frac{4M^2}{\ell^2}+\cdot\cdot\cdot)^2}{[2M(1+\frac{4M^2}{\ell^2}+\cdot\cdot\cdot)-M-\frac{16M^2}{\ell^2}(1+\frac{4M^2}{\ell^2}+\cdot\cdot\cdot)^3]}
[1\cr&&+\beta\{m^2+\frac{(1+\frac{4M^2}{\ell^2}+\cdot\cdot\cdot)}{[2M(1+\frac{4M^2}{\ell^2}+\cdot\cdot\cdot)-M-\frac{16M^2}{\ell^2}
(1+\frac{4M^2}{\ell^2}+\cdot\cdot\cdot)^3]}\}],
\end{eqnarray}

where $+(-)$ indicate the outgoing (ingoing) solution. Neglecting the terms $\leq M^{5}$ and applying WKB approximation, the tunneling rate of the fermions crossing the black hole event horizon is recovered as

\begin{eqnarray}
T=\frac{(1-\frac{8M^2}{\ell^2})}{8\pi M\omega(1+\frac{8M^2}{\ell^2})(1+\beta K)} =\frac{T_{0}}{(1+\beta K)},
\end{eqnarray}
\begin{figure}[ht]
  \centering
  \includegraphics[width=110mm]{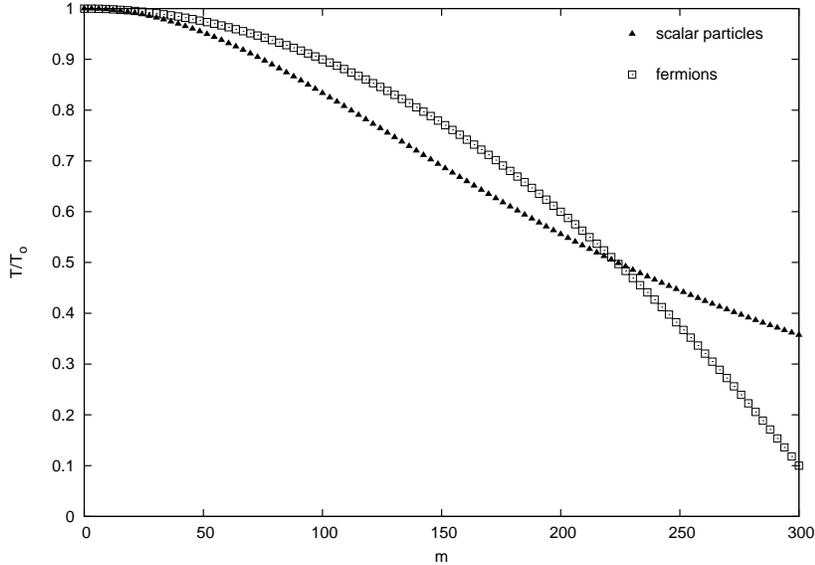}\\
  \caption{ Variations of the corrected Hawking temperatures with the masses of the emitted scalar particles and fermions (in arbitrary units) for SdS black hole}
% \caption{}\label{}
\end{figure}
where $K=m^{2}+\frac{4\omega^2(1+12M^2\ell^{-2})(1-8M^2\ell^{-2})}{(1-16M^4\ell^{-2})(1+8M^2\ell^{-2})}$. From above equation, $T_{0}=\frac{(1-8M^2\ell^{-2})}{8\pi M(1+8M^2\ell^{-2})}$ is the original Hawking temperature of SdS black hole in the absence of quantum gravity effect. Eq. (38) indicates that the corrected Hawking temperature is lower than the standard one. Due to quantum gravity effect, a small correction to the Hawking temperature is produced during black hole evaporation. It shows that the quantum gravity effects slow down the increase of the Hawking radiation temperature. The corrected Hawking temperature not only depends on the black hole mass but also on the mass and energy of the emitted fermion.

\begin{center}
 4. {\bf Discussion and Conclusion}
 \setcounter {equation}{0}
\renewcommand {\theequation}{4.\arabic{equation}}
 \end{center}

The corrections to the Hawking temperatures of the scalar particles and the fermions for the SdS black hole are studied using the generalized Klein-Gordon equation and the generalized Dirac equation repectively. The variations of the corrected Hawking temperatures with the mass's of the scalar particles and fermions are as shown in Fig. 1. For the scalar particles by direct calculation, the corrected Hawking temperature depends on the mass of the black hole and mass of the emitted particle only but not explicitly on the  energy of the emitted particle. Neglecting a small term in the roots of biquadratic equation and WKB approximation, the modified temperature depends not only on the mass of the black hole but also on the energy of the emitted particle. The tunneling rate near the black hole event horizon is related to the Bekenstein-Hawking entropy and a small term $(1+2\beta m^2)$. If $\beta=0$ and $\ell\longrightarrow0$, the SdS black hole hole becomes Schwarzschild black hole. In such case, the locations of the black hole event horizon of the Schwarzschild black hole before and after the emission of the particle with energy $\omega$ are $r_i=2M$ and $r_j=2(M-\omega)$ respectively.
From Eq. (21), the tunneling rate of the Schwarzschild black hole can be expressed as
\begin{eqnarray*}
\Gamma\sim exp[\Delta S_{BH}]=exp[-8\pi\omega(M-\frac{\omega}{2})].
\end{eqnarray*}
Our result is fully consistent with the result of Parikh and Wilczek [3].

In both the cases of tunneling of scalar particles and fermions, the corrected Hawking temperatures lower the original Hawking temperature showing that the quantum gravity effects slow down the rise of Hawking temperature of SdS black hole.

\end{document}